\documentclass[prl,twocolumn]{revtex4-1}
\usepackage{amsmath}
\usepackage{amssymb}
\usepackage{array}
\usepackage{mathrsfs}
\usepackage{tabularx}
\usepackage{graphicx}
\usepackage{dcolumn}
\usepackage{bm}
\usepackage{graphicx}
\usepackage{color}
\usepackage{booktabs}
\usepackage{multirow}
\usepackage{diagbox}

\begin{document}
\title{Measurement-device-independent quantum key distribution with insecure sources}
\author{Hua-Jian Ding$^{1,2,3}$}
\author{Xing-Yu Zhou$^{1,2,3}$}
\author{Chun-Hui Zhang$^{1,2,3}$}
\author{Jian Li$^{1,2,3}$}
\author{Qin Wang$^{1,2,3}$}
\email{qinw@njupt.edu.cn}

\affiliation{$^{1}$Institute of quantum information and technology, Nanjing University of Posts and Telecommunications, Nanjing 210003, China.}
\affiliation{$^{2}$"Broadband Wireless Communication and Sensor Network Technology" Key Lab of Ministry of Education, NUPT, Nanjing 210003, China.}
\affiliation{$^{3}$"Telecommunication and Networks" National Engineering Research Center, NUPT, Nanjing 210003, China.}

\begin{abstract}
Measurement-device-independent quantum key distribution (MDI-QKD) can eliminate all detector side-channel loopholes and has shown excellent performance in long-distance secret keys sharing. Conventional security proofs, however, require additional assumptions on sources and that can be compromised through uncharacterized side channels in practice. Here, we present a general formalism based on reference technique to prove the security of MDI-QKD against any possible sources' imperfection and/or side channels. With this formalism, we investigate the asymptotic performance of single-photon sources without any extra assumptions on the state preparations. Our results highlight the importance of transmitters' security.

\end{abstract}

\maketitle
\section{Introduction}
Quantum key distribution (QKD) can allow two legitimate users, Alice and Bob, to generate secret keys with information-theoretic security even in the presence of eavesdropper, Eve, who has unlimited computation powers. Since the first protocol, called BB84, is proposed by Bennett and Brassard in 1984 \cite{BB84}, QKD has achieved rapid developments theoretically and experimentally \cite{Xu, Pirandola}. In principle, QKD promises unconditional security based on quantum laws \cite{Shor, Lo, Mayers1} and enables permanent protection of confidential data when combined with Vernam's one-time pad cipher. In practice, however, realistic implementations would open security loopholes at the level of devices. These could be identified and exploited by Eve to enforce specific hacking and side-channel attacks \cite{Fung, Lydersen, Qi, Makarov}.

One way that can resist all side-channel attacks is fully device-independent (DI) QKD. 
However, DI-QKD is greatly challenging to realize in that it requires perfectly efficient detection efficiency and no information leakage from the measurement units.
As a compromise, a more practical strategy is proposed, namely measurement-device-independent (MDI) QKD \cite{MDI1,MDI2}. MDI-QKD is easy to implement with current technology and has been widely demonstrated \cite{EMDI2,EMDI4,EMDI5,EMDI6,EMDI7}. In terms of security, MDI-QKD can remove all potential detector side channels, but still makes additional assumptions on transmitters. To be precise, in a typical MDI-QKD system, Alice and Bob prepare almost perfect states from their fully protected laboratory \cite{MDI1}.
However, such premise on sources can be compromised through uncharacterized side channels, say, state preparation flaws (SPFs) \cite{GLLP}, information leakage \cite{Gisin}, and classical correlations between the generated pulses \cite{5GHz,Yoshino}, et al.
At present, there exist solutions for some security vulnerabilities. For example, SPFs have been efficiently treated with the loss-tolerant (LT) method \cite{LT,LTBB84,LTMDI} and the so-called uncharacterized qubit sources \cite{Yin1,Yin2}. Moreover, the issue of information leakage from users' internal settings has also been studied in Refs. \cite{THA1,THA2,GLT}. Lastly, the pulse correlations among emitted signals have been incorporated in recent works \cite{Yoshino,RT1}. Remarkably, the reference technique (RT) introduced in Ref. \cite{RT1} is general to accommodate various other side channels. Inspired by the results of RT, we combine it with MDI-QKD to guarantee practical security against both sources and detection side channels. For this, we consider some reference states and bound the maximum deviation between the probabilities associated with them and those associated with the actual emitted states. In particular, we evaluate the performance of the protocol with single-photon sources.

\section{Protocol Description}
\begin{figure}
	\centering
	\includegraphics[width=\linewidth,height=4cm]{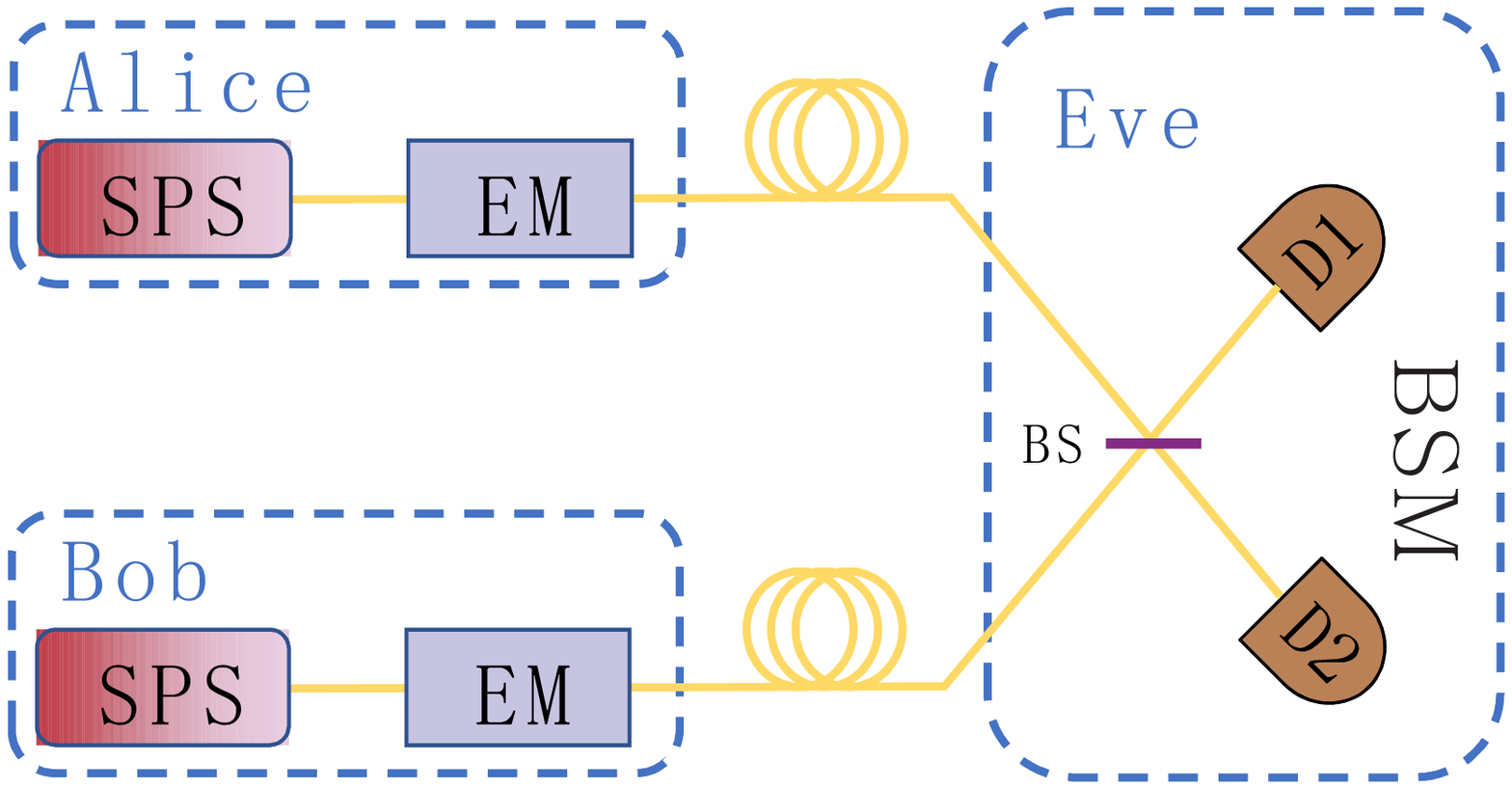}
	\caption{Schematic diagram of MDI-QKD with time-bin phase encoding. SPS: single-photon source; EM: encoding modulator; BS: beam splitter; BSM: Bell state measurement by using two detectors, D1 and D2.}
	\label{fig-1}
\end{figure}

For simplicity, we assume that there are no side channels in the following description. Figure \ref{fig-1} shows a typical MDI-QKD setup.  
\\
1. In each round, Alice (Bob) wants to generate the state ${\left| {{\varphi _{j_\alpha}}} \right\rangle _a}$ (${\left| {{\varphi _{s_\beta}}} \right\rangle _b}$), where $j,s  \in \left\{ {0,1} \right\}$ and $\alpha , \beta  \in \left\{ {Z,X} \right\}$ are their bit value and basis choices, respectively. As in the LT analysis \cite{LT}, they only select $j_\alpha , s_\beta  \in \left\{ {0_Z,1_Z,0_X} \right\}$. The states are then send out to an untrusted relay Eve via quantum channels.
\\
2. If Eve is honest, he performs a Bell state measurement (BSM) that projects the incoming signals into Bell state. Next, he announces the results of BSM. For simplicity, the discussion below only considers one Bell state: $\left| {{\psi ^ - }} \right\rangle$. 
\\
3. Alice and Bob keep the data that corresponds to the successful instances and discard the rest, regardless of whether they employ the same or different bases. Next, say Bob flips his data to correctly correlate them with those of Alice. 
\\
4. Alice and Bob reveal part of their sifted keys to estimate both the bit and the phase error rates. Finally, they perform error correction and privacy amplification to extract secret key strings.

In fact, for each particular round of the protocol, the emitted joint states are actually in the form 
\begin{equation}
{\left| {{{\Phi}_{j_\alpha,s_\beta}}} \right\rangle _T} = \sqrt {1 - {\varepsilon_{j_\alpha,s_\beta}}} {\left| {{\phi _{j_\alpha,s_\beta}}} \right\rangle _T} + \sqrt {{\varepsilon_{j_\alpha,s_\beta}}} {\left| {\phi _{j_\alpha,s_\beta}^ \bot } \right\rangle _T} ,	
\label{eq-1}
\end{equation} 
where $T:=abE$, which include Alice's (Bob's) transmitted system $a$ ($b$) and Eve's system $E$.  $\varepsilon_{j_\alpha,s_\beta}$ is a non-negative real number that satisfys $0\leqslant \varepsilon_{j_\alpha,s_\beta} \leqslant 1$. ${\left| {{\phi _{j_\alpha,s_\beta}}} \right\rangle _T}: = {\left| {{\varphi _{j_\alpha}}} \right\rangle _a}{\left| {{\varphi _{s_\beta}}} \right\rangle _b}{\left| \tau  \right\rangle _E}$ with ${\left| \tau  \right\rangle _E}$ being a state that does not contain any information about Alice’s and Bob’s current round selections, and ${\left| {\phi _{j_\alpha,s_\beta}^ \bot } \right\rangle _T}$ is an unknown side-channels state orthogonal to ${\left| {\phi _{j_\alpha,s_\beta}} \right\rangle _T}$. 
Importantly, any potential side channels from transmitters can be characterized with Eq. (\ref{eq-1}), which thus represents the most general description of the emitted states. This have been detailedly substantiated in Ref. \cite{RT2}.

The asymptotic key rate for single-photon sources is given by  
\begin{equation}
R \ge {Y_{ZZ}}\left[ {1 - h\left( {{e_{XX}}} \right) - {f_{EC}} h\left( {{e_{ZZ}}} \right)} \right],
\label{eq-2}
\end{equation} 
where $Y_{ZZ}$ and $e_{ZZ}$ are  the yield and bit error rate in $ZZ$ basis, respectively, and can be directly obtained from experiment. The function of $h(\cdot)$ is the binary entropy function and $f_{EC}(=1.16)$ is the error correction efficiency. The term $e_{XX}$ is the phase error rate, which is an essential parameter to be estimated. For this, we use the complementary augment introduced by Koashi \cite{Koashi},  where an equivalent virtual protocol is created.

In the virtual protocol, from Eve's perspective, Alice and Bob first prepare the following state in the $ZZ$ basis:
\begin{equation}
{\left| {{{\Phi }^{{\rm{vir}}}}} \right\rangle _{ABT}} = \frac{1}{2}\mathop \sum \limits_{j,s = 0,1} {\left| {j_Z,s_Z} \right\rangle _{AB}}{\left| {{{\Phi }_{j_Z,s_Z}}} \right\rangle _T} ,
\label{eq-3}
\end{equation}
with $ \{ 0_Z,1_Z \} $ being the computational basis for ancillary systems $A$ and $B$, and subsequently they send the system $T$ to Eve. We then define the bit error rate as
\begin{equation}
{e_{ZZ}} = \frac{{Y_{0_Z,0_Z}^{(ZZ)} + Y_{1_Z,1_Z}^{(ZZ)}}}{{Y_{0_Z,0_Z}^{(ZZ)} + Y_{1_Z,0_Z}^{(ZZ)} + Y_{0_Z,1_Z}^{(ZZ)} + Y_{1_Z,1_Z}^{(ZZ)}}},
\label{eq-4}
\end{equation}
where the yield $ Y_{j_Z,s_Z}^{(ZZ)} $ is the joint probability that Eve declare a successful BSM when Alice and Bob first prepare the state ${\left| {{{\Phi }^{{\rm{vir}}}}} \right\rangle _{ABT}}$ and Alice (Bob) obtains the bit value $j$ ($s$) by measuring the system $A$ ($B$) in the $Z$ basis. Note that the superscripts $ZZ$ denote the bases employed in state preparation, while the subscripts represent the bases used in local measurement. 
For brevity of notation, we shall omit the bases superscript or mode subscript, unless otherwise needed. 
Similarly, the phase error rate is defined as 
\begin{equation}
{e_{XX}} = \frac{{Y_{0_X,0_X}^{(ZZ)\rm{vir}} + Y_{1_X,1_X}^{(ZZ)\rm{vir}}}} {{Y_{0_X,0_X}^{(ZZ)\rm{vir}} + Y_{1_X,0_X}^{(ZZ)\rm{vir}} + Y_{0_X,1_X}^{(ZZ)\rm{vir}} + Y_{1_X,1_X}^{(ZZ)\rm{vir}}}},
\label{eq-5}
\end{equation} 
where ${Y_{s_X,j_X}^{(ZZ)\rm{vir}}}$ is the joint probability that Alice (Bob) obtain bit value $s$ ($j$) in the virtual X-basis measurement on system $A$ ($B$) given the state preparation ${\left| {{{\Phi }^{{\rm{vir}}}}} \right\rangle _{ABT}}$ and also Eve declares a successful BSM. The phase error rate corresponds to the bit error in the virtual protocol. In addition, we have that the denominator of ${e_{XX}}$ in Eq. (\ref{eq-5}) is equal to $ \sum\limits_{j,s = {0,1}} {Y_{j_Z,s_Z}} =: \zeta _{obs} $. Therefore, we only need to calculate the numerator $\Omega := {Y_{0_X,0_X}^{\rm{vir}}} + {Y_{1_X,1_X}^{\rm{vir}}}$ for $e_{XX}$. Note that after Alice and Bob complete the virtual X-basis measurement, they send Eve the unnormalized states:
\begin{equation}
\hat \Theta _{j,s}^{\rm{vir}} = Tr_{AB} \left[ {\left| {j_X,s_X} \right\rangle {{\left\langle {j_X,s_X} \right|}_{AB}} \otimes {\mathbb{1}_T}\left| {{{{\Phi }}^{{\rm{vir}}}}} \right\rangle {{\left\langle {{{{\Phi }}^{{\rm{vir}}}}} \right|}_{ABT}}} \right],
\label{eq-6}
\end{equation}
where $ Tr_{AB} $ is the partial trace over ancillary systems $AB$. We write the normalized version as $\Theta _{j,s}^{\rm{vir}} = \hat \Theta _{j,s}^{\rm{vir}} / Tr \left( {\hat \Theta _{j,s}^{\rm{vir}}} \right)$.



To find the unknown quantity $\Omega$, we employ the RT method \cite{RT1}, namely considering some reference states that close to their respective actual states. These reference states, in principle, can be chosen freely, but they should be selected in a way that it is easy to derive a relationship among the probabilities associated with them. For this, as an example, we select the reference states to be $ \left\{ {\left| \varphi _{0_Z} \right \rangle , \left| \varphi _{1_Z} \right \rangle , \left| \varphi _{0_X} \right \rangle  } \right\} $ for each user, which are defined as 
\begin{equation}
\begin{array}{l}
\left| {{\varphi _{0_Z}}} \right\rangle  = \cos \left( {\frac{{{\delta _1}}}{2}} \right)\left| {{0_Z}} \right\rangle  + \sin \left( {\frac{{{\delta _1}}}{2}} \right)\left| {{1_Z}} \right\rangle ,\\
\left| {{\varphi _{1_Z}}} \right\rangle  = \sin \left( {\frac{{{\delta _2}}}{2}} \right)\left| {{0_Z}} \right\rangle  + \cos \left( {\frac{{{\delta _2}}}{2}} \right)\left| {{1_Z}} \right\rangle ,\\
\left| {{\varphi _{0_X}}} \right\rangle  = \sin \left( {\frac{\pi }{4} + \frac{{{\delta _3}}}{2}} \right)\left| {{0_Z}} \right\rangle  + \cos \left( {\frac{\pi }{4} + \frac{{{\delta _3}}}{2}} \right)\left| {{1_Z}} \right\rangle ,
\end{array}
\label{eq-7}
\end{equation}
where $ \delta _i (i=1,2,3)$ denote the deviations of the phase modulation from the intended values due to encoding modulators. 
We emphasize that these reference states are never perpared in actual protocol but serve as mathematical tool for parameter estimation.

From the definitions of $ \left| {{{\Phi }^{{\rm{vir}}}}} \right\rangle $, $\Theta _{j,s}^{\rm{vir}}$ and ${{Y}_{s_X,j_X}^{\rm{vir}}}$, we can define analogous states and probabilities $ \left| {{{\Psi }^{{\rm{vir}}}}} \right\rangle $, $\Theta _{j,s}^{\rm{vir}|ref}$ and ${{Y}_{s_X,j_X}^{\rm{vir}|ref}}$ for reference states. In particular, the yields ${{Y}_{s_X,j_X}^{\rm{vir}|ref}}$ are defined as 
\begin{equation}
{Y}_{s_X,j_X}^{{\rm{vir|ref}}} = p_{j,s}^{{\rm{vir|ref}}}Tr\left[ {{{\hat M}_{{\psi ^ - }}}\Theta _{j,s}^{{\rm{vir|ref}}}} \right] ,
\label{eq-8}
\end{equation}
where $ {{{\hat M}_{{\psi ^ - }}}} $ corresponds to the successful announcement of Eve's BSM, and $ p_{j,s}^{{\rm{vir|ref}}} = Tr \left[ {\Theta _{j,s}^{{\rm{vir|ref}}}} \right] $. Again, we define an analogous quantity $\Omega _{\rm{ref}} := {{Y}_{0_X,0_X}^ {\rm{vir|ref}}} + {{Y}_{1_X,1_X}^{\rm{vir|ref}}}$ for the reference states, and then evaluate the deviation between the probabilities associated with the reference states and those associated with the actual states. Following the analysis of Ref. \cite{RT1}, this deviation is quantified by
\begin{equation}
\begin{array}{l}
{G^{\rm{L}}}\left( {\left\langle A \right|\hat M\left| A \right\rangle ,\left| {\left\langle {A}
		\mathrel{\left | {\vphantom {A R}}
			\right. \kern-\nulldelimiterspace}
		{R} \right\rangle } \right|} \right) \le \left\langle R \right|\hat M\left| R \right\rangle  \le \\
{G^{\rm{U}}}\left( {\left\langle A \right|\hat M\left| A \right\rangle ,\left| {\left\langle {A}
		\mathrel{\left | {\vphantom {A R}}
			\right. \kern-\nulldelimiterspace}
		{R} \right\rangle } \right|} \right)
\end{array}
\label{eq-9}
\end{equation}
where $\left| A \right\rangle $ and $\left| R \right\rangle $ are normalized pure state associated with the actual and reference states, respectively. ${\hat M}$ is any non-negative bounded operator such that ${0 \le \hat M \le 1}$, and we define $ \hat M = \left( {\left| {{0_x},{0_x}} \right\rangle \left\langle {{0_x},{0_x}} \right| + \left| {{1_x},{1_x}} \right\rangle \left\langle {{1_x},{1_x}} \right|} \right) \otimes {{{\hat M}_{{\psi ^ - }}}} $. By applying \textit{Cauchy-Schwarz inequality} to the vectors $ \sqrt {\hat N} \left| A \right\rangle $ and $ \sqrt {\hat N} \left| R \right\rangle $, with one $ \hat N = \hat M $ and another $ \hat N = \hat I - \hat M $, we can get the functions ${G^{\rm{L}}}\left( {x,y} \right)$ and ${G^{\rm{U}}}\left( {x,y} \right)$ as follows
{\footnotesize
	\begin{equation}
	{G^{\rm{L}}}\left( {x,y} \right) = \left\{ {\begin{array}{*{20}{c}}
		0&{x < 1 - {y^2}}\\
		{x + \left( {1 - {y^2}} \right)(1 - 2x) - 2y\sqrt {\left( {1 - {y^2}} \right)x(1 - x)} }&{x \ge 1 - {y^2}}
		\end{array}} \right.
	\label{eq-10}
	\end{equation}
}
and
{\footnotesize
	\begin{equation}
	{G^{\rm{U}}}\left( {x,y} \right) = \left\{ {\begin{array}{*{20}{c}}
		{x + \left( {1 - {y^2}} \right)(1 - 2x) + 2y\sqrt {\left( {1 - {y^2}} \right)x(1 - x)} }&{x \le {y^2}}\\
		1&{x > {y^2}}
		\end{array}} \right. .
	\label{eq-11}
	\end{equation} 
}Note that $ {-G^{\rm{L}}}\left( {x,y} \right) $ and $ {G^{\rm{U}}}\left( {x,y} \right) $ are concave with respect to $ 0 \le x \le 1 $ for any fixed $ 0 \le y \le 1 $, and $ {\partial _y} {G^{\rm{L}}}\left( {x,y} \right) \geq 0 $ and $ {\partial _y} {G^{\rm{U}}}\left( {x,y} \right) \leqslant 0 $ hold. And then, since $ \Omega = \left\langle {{\Phi ^{{\rm{vir}}}}} \right|\hat M\left| {{\Phi ^{{\rm{vir}}}}} \right\rangle $ and $ {\Omega _{{\rm{ref}}}} = \left\langle {{\Psi ^{{\rm{vir}}}}} \right|\hat M\left| {{\Psi ^{{\rm{vir}}}}} \right\rangle $, we can employ Eq. (\ref{eq-9}) to get an upper bound on $\Omega$:
\begin{equation}
\Omega  \le {G^{\rm{U}}}\left( {{\Omega _{{\rm{ref\;}}}},{\delta _{{\rm{vir\;}}}}} \right) \le {G^{\rm{U}}}\left( {\Omega _{{\rm{ref\;}}}^{\rm{U}},\delta _{{\rm{vir\;}}}^{\rm{L}}} \right) = :{\Omega ^{\rm{U}}} ,
\label{eq-12}
\end{equation} 
where $ {\Omega _{{\rm{ref\;}}}^{\rm{U}}} $ is an upper bound on $ {{\Omega _{{\rm{ref\;}}}}} $, 
and $\delta _{{\rm{vir\;}}}^{\rm{L}} = \frac{1}{4} \mathop \sum \limits_{j,s = 0,1} \sqrt {1 - {\varepsilon _{{j_Z},{s_Z}}}} $ is a lower bound on $\delta _{{\rm{vir\;}}}:= \left| {\left\langle {{{\Psi ^{{\rm{vir}}}}}} \mathrel{\left | {\vphantom {{{\Psi ^{{\rm{vir}}}}} {{\Phi ^{{\rm{vir}}}}}}} \right. \kern-\nulldelimiterspace} {{{\Phi ^{{\rm{vir}}}}}} \right\rangle } \right| $. Importantly, $ {\Omega _{{\rm{ref\;}}}^{\rm{U}}} $ can be bounded from all observables $ {Y}_{{j_\alpha },{s_\beta }} $ of the actual states and from the square root fidelities of $ \left| {\left\langle {{{\Psi _{{j_\alpha },s_\beta }}}} \mathrel{\left | {\vphantom {{{\Psi _{{j_\alpha },s_\beta }}} {{\Phi _{j_\alpha ,s_\beta }}}}} \right. \kern-\nulldelimiterspace} {{{\Phi _{j_\alpha ,s_\beta }}}} \right\rangle } \right| $, with ${\left| {{\Psi _{j_\alpha ,s_\beta }}} \right\rangle _T} = {\left| {{\varphi _{j_\alpha }}} \right\rangle _a}{\left| {{\varphi _{s_\beta }}} \right\rangle _b}{\left| \tau  \right\rangle _E}$. 

Below, we show how to obtain $ {\Omega _{{\rm{ref\;}}}^{\rm{U}}} $ in detail. According to the LT method \cite{LT}, the virtual states $ \Theta _{j,s}^{{\rm{vir}}\mid {\rm{ref}}} $ can be expressed as a linear combination of the \textit{Pauli operators} $ \left\{ {{\sigma _I},{\sigma _X},{\sigma _Z}} \right\} $:

\begin{equation}
\Theta _{j,s}^{{\rm{vir}}\mid {\rm{ref}}} = \frac{1}{4} \sum \limits_{l,l'} S_{l,l'}^{j,s|{\rm{vir}}}\sigma _l \otimes \sigma _{l'} ,
\label{eq-13}
\end{equation}
where $ S_{l,l'}^{j,s|{\rm{vir}}} $ with $ l,l' \in \left\{ {I,X,Z} \right\} $ are the coefficients of the Bloch vector. Define the transmission rate of $ \sigma _l \otimes \sigma _{l'} $ as 
\begin{equation}
{q_{l,l'}} = \frac{1}{4}Tr\left[ {\hat M_{{\psi ^ - }} \sigma _l \otimes \sigma _{l'}} \right] ,
\label{eq-14}
\end{equation}
and then combine it with Eqs. (\ref{eq-8}) and (\ref{eq-13}), the transmission rate $ Y_{j,s}^{{\rm{vir|ref\;}}} $ can be rewritten as
\begin{equation}
Y_{j,s}^{{\rm{vir|ref\;}}} = p_{j,s}^{{\rm{vir|ref}}} \mathop \sum \limits_{l,l'} S_{l,l'}^{j,s{\rm{|vir\;}}} {q_{l,l'}} .
\label{eq-15}
\end{equation} 
With this notation, we can concisely write the matrix equation 
\begin{equation}
{\Omega _{{\rm{ref}}}} =  {{{\rm{\textbf{P}}}^{{\rm{vir}}}}}   {{\rm{\textbf{S}}}^{{\rm{vir}}}}{\rm{\textbf{q}}} ,
\label{eq-16}
\end{equation}
where $ {{{\rm{\textbf{P}}}^{{\rm{vir}}}}} =  \left[ {p_{0,0}^{{\rm{vir|ref}}}, p_{1,1}^{{\rm{vir|ref}}}} \right] $, $ {{\rm{\textbf{S}}}^{{\rm{vir}}}} $ is a $ 2\times9 $ matrix containing the coefficients $ S_{l,l'}^{0,0{\rm{|vir\;}}} $ ($ S_{l,l'}^{1,1{\rm{|vir\;}}} $) in its first (second) row, and {\rm{\textbf{q}}} is a column vector containing the quantities $ {q_{l,l'}} $. Similarly, we have  
\begin{equation}
{\rm{\textbf{Y}}^{{\rm{ref}}}}{\rm{ = \textbf{Sq}}} ,
\label{eq-17}
\end{equation}
wher $ {\rm{\textbf{Y}}^{{\rm{ref}}}} $ is a column vector containing the yields $ {Y}_{{j_\alpha },{s_\beta }}^{{\rm{ref\;}}} $, and $ \textbf{S} $ is a $ 9\times9 $ matrix containing the Bloch coefficients of the reference states $ \left| {{\Psi _{j\alpha ,s\beta }}} \right\rangle $ in its rows. Then, by combining Eqs. (\ref{eq-16}) and (\ref{eq-17}), we can conveniently get 
\begin{equation}
{\Omega _{{\rm{ref}}}} =  {{{\rm{\textbf{P}}}^{{\rm{vir}}}}}  {{\rm{\textbf{S}}}^{{\rm{vir}}}}{{\rm{\textbf{S}}}^{ - 1}}{\rm{\textbf{Y}}^{{\rm{ref}}}} . 
\label{eq-18}
\end{equation}
Note that once the reference states are selected, these associated matrices $ {\rm{\textbf{P}}}^{{\rm{vir}}} $, $ {\rm{\textbf{S}}}^{{\rm{vir}}} $ and $ {\rm{\textbf{S}}} $ are determined. If we further define $ {{\bf{f}}_{{\rm{obj}}}}: = {{{\bf{P}}^{{\rm{vir}}}}} {{\bf{S}}^{{\rm{vir}}}} {{\bf{S}}^{ - 1}} $ and bound each term in Eq. (\ref{eq-18}) separately, we can obtain the upper bound on $ {\Omega _{{\rm{ref\;}}}} $:
\begin{align}
{\Omega _{{\rm{ref}}}} =& {{{\bf{f}}_{{\rm{obj}}}}{{\bf{Y}}^{{\rm{ref}}}} = \sum\limits_{{j_\alpha },{s_\beta }} {{f_{{j_\alpha },{s_\beta }}} {Y}_{{j_\alpha },{s_\beta }}^{{\rm{ref}}}} } \\ \nonumber
\le& {\mathop \sum \limits_{{j_\alpha },{s_\beta }\mid {f_{{j_\alpha },{s_\beta }}} > 0} {f_{{j_\alpha },{s_\beta }}}{G^{\rm{U}}}\left( {{{Y}_{{j_\alpha },{s_\beta }}},\delta _{{j_\alpha },{s_\beta }}^L} \right)} \\ \nonumber
{}& { + \mathop \sum \limits_{{j_\alpha },{s_\beta }\mid {f_{{j_\alpha },{s_\beta }}} < 0} {f_{{j_\alpha },{s_\beta }}}{G^{\rm{L}}}\left( {{{Y}_{{j_\alpha },{s_\beta }}},\delta _{{j_\alpha },{s_\beta }}^L} \right)} \\ \nonumber
=& :{\Omega _{{\rm{ref}}}^{\rm{U}}} 
\label{eq-19}
\end{align}
with $ {f_{{j_\alpha },{s_\beta }}} $ being the elements of the vector $ {{\bf{f}}_{{\rm{obj}}}} $. The terms $ {\delta _{{j_\alpha },{s_\beta }}^{\rm{L}}} $ are lower bound on $ \left| {\left\langle {{{\Psi _{{j_\alpha },s_\beta }}}}\mathrel{\left | {\vphantom {{{\Psi _{{j_\alpha },s_\beta }}} {{\Phi _{j_\alpha ,s_\beta }}}}} \right. \kern-\nulldelimiterspace}{{{\Phi _{j_\alpha ,s_\beta }}}} \right\rangle } \right| $, and roughly set as $ \delta _{{j_\alpha },{s_\beta }}^{\rm{L}} = \sqrt {1 - {\varepsilon _{{j_\alpha},{s_\beta}}}} $. Notably, in the absence of observed statistics $ {{Y_{{j_\alpha },{s_\beta }}}} $, we instead use the $ {Y}_{{j_\alpha },{s_\beta }}^{{\rm{ref\;}}} $ according to a typical channel model \cite{WangQ} for numerical simulation. Finally, the phase error rate can be estimated by $ e_{_{XX}} = {\Omega ^{\rm{U}}}/ \zeta _{obs} $.

\begin{figure}
	\centering
	\includegraphics[width=\linewidth,height=\columnwidth]{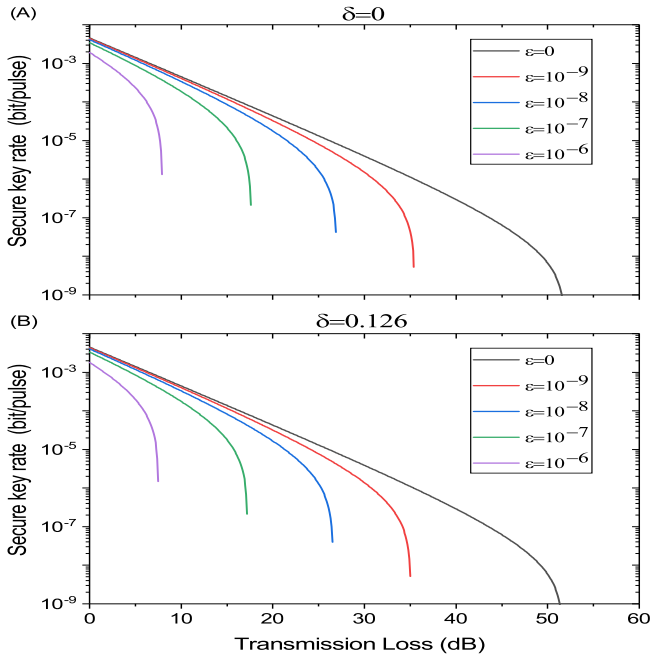}
	\caption{Secret key rate versus transmission loss (dB) in the presence of side channels.  (\textbf{A})There are no SPFs. (\textbf{B}) When there exits small SPFs, the secret key rate is only slightly worse.}
	\label{fig-2}
\end{figure}

To show the performance of MDI-QKD in the presence of side channels, we now present the simulation results. The experimental parameters used are as follows: detection efficiency $ {\eta _d} = 14.5\% $ and dark-count probability $ p_d = 6.02\times10^{-6} $ of Eve's detectors, the intrinsic error rate $ e_d = 1.5\% $ \cite{MDI1}, and the probabilities for Alice and Bob select $Z$ basis are, for simplicity, $ {p_{{Z_A}}} = {p_{{Z_B}}} = \frac{2}{3} $. According to the experimantal data \cite{LTBB84,Honji}, we choose $ \delta_i =\delta =0 $ (0.126) for Eq. (\ref{eq-7}). Note that these SPFs from modulation require a characterization. Regarding side channels $ \varepsilon _{j_\alpha,s_\beta} $, unfortunately, there are no studies founded to fully describe it. Therefore, we select some values to evaluate this imperfection and, for simplicity, we set $ \varepsilon _{j_\alpha,s_\beta} = \varepsilon $ for all $ j_\alpha,s_\beta \in \left\{ {0_Z,1_Z,0_X} \right\} $. The main results are illustrated in Fig. \ref{fig-2}.  

As expected, the secret key rate sharply decreases when the side channels characterized by $\varepsilon$ increases. We note, however, that a positive key is still available when considering a large $ \varepsilon $. For instance, when $ \varepsilon = 10^{-6} $, Alice and Bob could generate a secret key over about 8-dB transmission loss. In addition, comparing panels (\textbf{A}) and (\textbf{B}) of Fig. \ref{fig-2}, the secret key rate corresponding to the case $ \delta=0$ and $ \delta =0.126 $ are almost overlap, which fully demonstrates the high tolerance against SPFs with channel loss of the LT method.

\begin{figure}
	\centering
	\includegraphics[width=\linewidth,height=5.4cm]{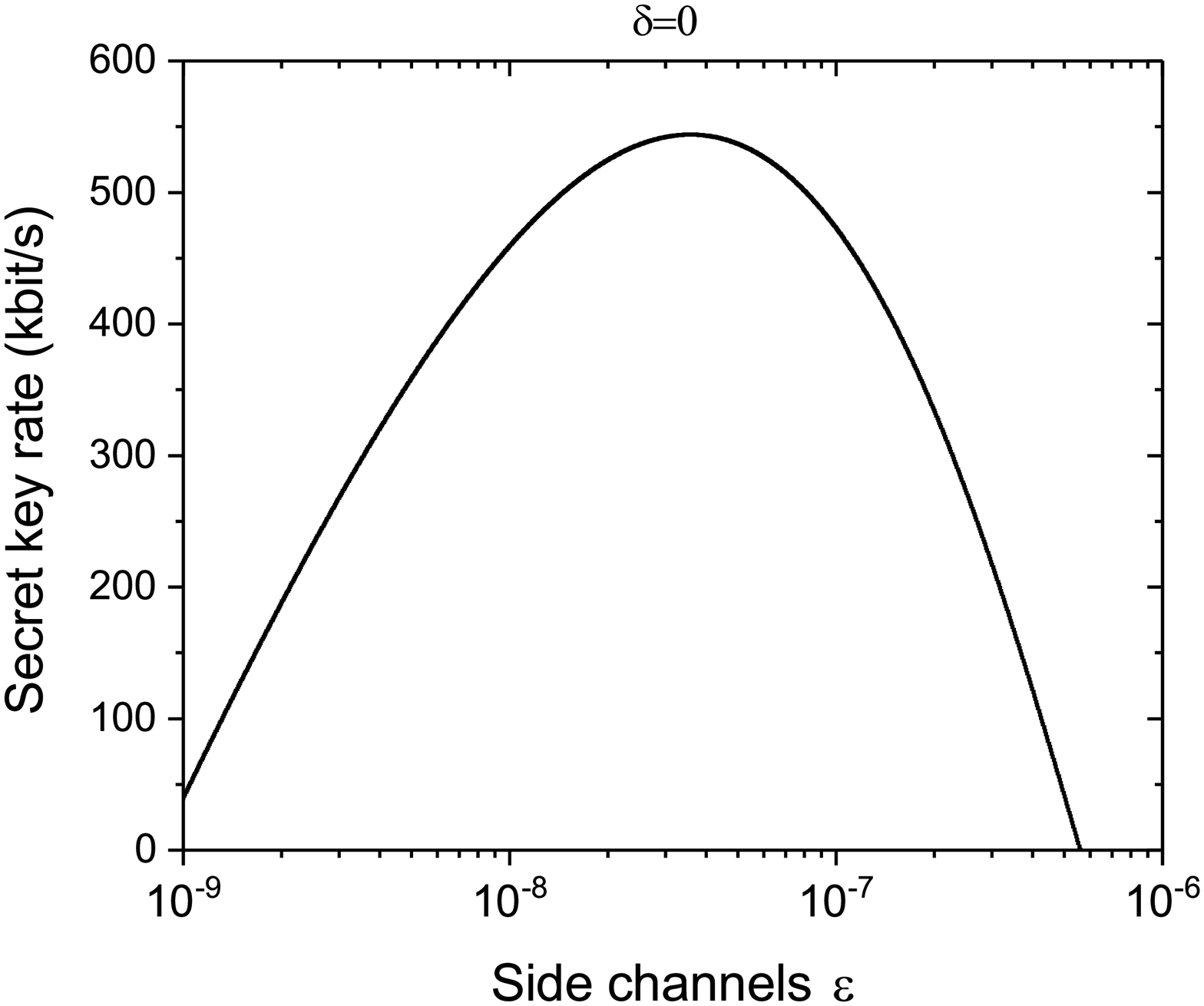}
	\caption{Secret key rate versus side channels. Here we assume that the side channels $ \varepsilon $ in lg is proportional to system frequency $f$, with $ \varepsilon \in \left[ {{10}^{ - 9},{10}^{ - 6}} \right] $ and $ f \in \left[ {0.1GHz, 4GHz}  \right] $. }
	\label{fig-3}
\end{figure}

Furthermore, comparing the purple and the black line of Fig. \ref{fig-2}, the secret key rates are quite different, which indicates that the side channels have a great influence on the key rate. The side channels, especially mode dependencies and pulse correlations, mostly occur in high-speed systems \cite{5GHz,Yoshino}. If we ignore them in security analysis, the achieved keys may not be secure. As shown in Fig. \ref{fig-3}, where we simply assume that lg($ \varepsilon $) is proportional to the system frequency $f$, the key rate reaches its maximum at some point, rather than increasing all the way as we would expect. Therefore, in the development of high-speed QKD systems, we need to strictly analyze the possible side channels to guarantee the security of shared keys.    

To conclude, we have introduced an MDI-QKD protocol that can accommodate any side channels in transmitters, thus closing the gap between theory and reality. 
The security analysis is achieved by introducing reference states and then bounding the maximum deviation between the probabilities associated with them and those of the actual states. 
Our results show that secret keys can be surely distributed with any side channel. In this regard, we believe that our work represents an important step towards constructing a truly secure QKD with realistic devices. The next steps for our protocol would be to incorporate coherent secure \cite{Azuma} and decoy-state methods \cite{Decoy2,Decoy3,RT3}, and complete the experimental characterization of $\varepsilon$ in practice.

\noindent\textbf{Funding.} National Key Research and Development Program of China (2018YFA0306400 and 2017YFA0304100); National Natural Science Foundation of China (NSFC) (12074194, 11774180 and U19A2075); Leading-edge Technology Program of Jiangsu Natural Science Foundation (BK20192001); Postgraduate Research \& Practice Innovation Program of Jiangsu Province (KYCX20\_0726).
\\
\\


\begin{thebibliography}{}
\bibitem{BB84} C. H. Bennett, and G. Brassard, in Proceedings of the IEEE International Conference on Computers, Systems and Signal Processing, Bangalore, India (1984), pp. 175-179.


\bibitem{Xu} F. H. Xu, X. F. Ma, Q. Zhang, H. K. Lo and J. W. Pan, "Secure quantum key distribution with realistic devices", Rev. Mod. Phys. \textbf{92}, 025002 (2020).

\bibitem{Pirandola}S. Pirandola, U. L. Andersen et al., "Advances in quantum cryptography", Adv. Opt. Photon. \textbf{12}, 1012 (2020).

\bibitem{Lo} H. K. Lo and H. F. Chau, "Unconditional Security of Quantum Key Distribution over Arbitrarily Long Distances", Science \textbf{283}.2050 (1999).

\bibitem{Shor}  P. W. Shor and J. Preskill, "Simple proof of security of the BB84 quantum key distribution protocol", Phys. Rev. Lett. \textbf{85}, 441 (2000).

\bibitem{Mayers1} D. Mayers, "Unconditional Security in Quantum Mechanics", J. ACM \textbf{48}, 351 (2001).

\bibitem{Makarov} V. Makarov, A. Anisimov, and J. Skaar, “Effects of detector efficiency mismatch on security of quantum cryptosystems”, Phys. Rev. A \textbf{74}, 022313 (2006).

\bibitem{Qi} B. Qi, C.-H. -F. Fung, H.-K. Lo, and X. Ma, “Time-shift attack in practical quantum cryptosystems”, Quantum Inf. Comput. \textbf{7}, 73 (2007).

\bibitem{Fung} C.-H. F. Fung, B. Qi, K. Tamaki, and H.-K. Lo, “Phase-remapping attack in practical quantum-key-distribution systems”, Phys. Rev. A \textbf{75}, 032314 (2007).

\bibitem{Lydersen} L. Lydersen, C. Wiechers, C. Wittmann, D. Elser, J. Skaar, and V. Makarov, “Hacking commercial quantum cryptography systems by tailored bright illumination”, Nat. Photonics. \textbf{4}, 686 (2010).

\bibitem {MDI1} H.-K. Lo, M. Curty, and B. Qi, “Measurement-device-independent quantum key distribution”, Phys. Rev. Lett. \textbf{108}, 130503 (2012).

\bibitem{MDI2} S. L. Braunstein and S. Pirandola, “Side-channel-free quantum key distribution”, Phys. Rev. Lett. \textbf{108}, 130502 (2012).

\bibitem{EMDI2} C. Wang, X. T. Song, Z. Q. Yin, S. Wang, W. Chen, C. M. Zhang, G. C. Guo, and Z. F. Han, "Phase-reference-free experiment of measurement-device-independent quantum key distribution", Phys. Rev. Lett. \textbf{115}, 160502 (2015).

\bibitem{EMDI4} L. C. Comandar, M. Lucamarini, B. Frhlich, J. F. Dynes, A.W. Sharpe, S.W. B. Tam, Z. L. Yuan, R. V. Penty, and A. J. Shields, "Quantum key distribution without detector vulnerabilities using optically seeded lasers", Nat. Photonics \textbf{10}, 312 (2016).

\bibitem{EMDI5} X. Y. Zhou, H. J. Ding, C. H. Zhang, J. Li, C. M. Zhang and Q. Wang, "Experimental three-state measurement-device-independent quantum key distribution with uncharacterized sources", Opt. Lett. \textbf{45}, 4176 (2020).

\bibitem{EMDI6} R. I. Woodward, Y. S. Lo, M. Pittaluga M. Minder, T. K. Paraïso, M. Lucamarini, Z. L. Yuan and A. J. Shields, "Gigahertz measurement-device-independent quantum key distribution using directly modulated lasers", Npj Quantum Inf. \textbf{7}, 58 (2021)

\bibitem{EMDI7} Y. P. Chen, J. Y. Liu, M. S. Sun, X. Y. Zhou, C. H. Zhang, J. Li, and Q. Wang, "Experimental measurement-device-independentquantum key distribution with the double-scanningmethod", arXiv:2105.09587 (2021).

\bibitem{GLLP} D. Gottesman, H.-K. Lo, N. Lütkenhaus, J. Preskill, "Security of quantum key distribution
with imperfect devices", Quantum Inf. Comput. \textbf{4}, 325–360 (2004).

\bibitem{Gisin} N. Gisin, S. Fasel, B. Kraus, H. Zbinden, G. Ribordy, "Trojan-horse attacks on quantum-key-distribution systems", Phys. Rev. A \textbf{73}, 022320 (2006).

\bibitem{Yoshino} K.-i. Yoshino, M. Fujiwara, K. Nakata, T. Sumiya, T. Sasaki, M. Takeoka, M. Sasaki, A. Tajima, M. Koashi, A. Tomita, "Quantum key distribution with an efficient countermeasure against correlated intensity fluctuations in optical pulses", Npj Quantum Inf. \textbf{4}, 8 (2018).

\bibitem{5GHz} F. Grünenfelder, F., Boaron, A., Rusca, D., Martin, A. and Zbinden, H. "Performance and security of 5 GHz repetition rate polarization-based quantum key distribution", Appl. Phys. Lett. \textbf{117}, 144003 (2020).

\bibitem{LT} K. Tamaki, M. Curty, G. Kato, H.-K. Lo, and K. Azuma, "Loss-tolerant quantum cryptography with imperfect sources", Phys. Rev. A \textbf{90}, 052314 (2014).

\bibitem{LTBB84} F. Xu, K. Wei, S. Sajeed, S. Kaiser, S. Sun, Z. Tang, L. Qian, V. Makarov, and H.-K. Lo, "Experimental quantum key distribution with source flaws", Phys. Rev. A \textbf{92}, 032305 (2015).

\bibitem{LTMDI} Z. Tang, K. Wei, O. Bedroya, L. Qian, and H. K. Lo, "Experimental measurement-device-independent quantum key distribution with imperfect sources", Phys. Rev. A \textbf{93}, 042308 (2016).

\bibitem{Yin1}Z. Q. Yin, C. H. F. Fung, X. Ma, C. M. Zhang, H. W. Li, W. Chen, S. Wang, G. C. Guo, and Z. F. Han, "Measurement-device-independent quantum key distribution with uncharacterized qubit sources", Phys. Rev. A \textbf{88}, 062322 (2013).

\bibitem{Yin2}Z. Q. Yin, C. H. F. Fung, X. Ma, C. M. Zhang, H. W. Li, W. Chen, S. Wang, G. C. Guo, and Z. F. Han, "Mismatched-basis statistics enable quantum key distribution with uncharacterized qubit sources", Phys. Rev. A \textbf{90}, 052319 (2014).

\bibitem{THA1} M. Lucamarini, I. Choi, M. B. Ward, J. . F. Dynes, Z. L. Yuan, A. J. Shields, "Practical security bounds against the trojan-horse attack in quantum key distribution", Phys. Rev. X \textbf{5}, 031030 (2015).

\bibitem{THA2} K. Tamaki, M. Curty, and M. Lucamarini, "Decoy-state quantum key distribution with a leaky source", New J. Phys. \textbf{18}, 065008 (2016).

\bibitem{Side-channel-free} X. B. Wang, X. L. Hu and Z. W. Yu, "Practical Long-Distance Side-Channel-Free Quantum Key Distribution", Phys. Rev. Appl. \textbf{12}, 054034 (2019).

\bibitem{GLT} M. Pereira, M. Curty, and K. Tamaki, "Quantum key distribution with flawed and leaky sources", Npj Quantum Inf. \textbf{5}, 1 (2019).

\bibitem{RT1} M. Pereira, G. Kato, A. Mizutani, M. Curty, and K. Tamaki, "Quantum key distribution with correlated sources", Sci. Adv. \textbf{6}, eaaz4487 (2020).

\bibitem{RT2} Á. Navarrete, M. Pereira, M. Curty, and K. Tamaki, "Practical Quantum Key Distribution That is Secure Against Side Channels",  Phys. Rev. Appl. \textbf{15}, 034072 (2021).

\bibitem{Koashi} M. Koashi, "Simple security proof of quantum key distribution based on complementarity", New J. Phys. \textbf{11}, 045018 (2009).

\bibitem{WangQ} Q. Wang, X.B. Wang, "Simulating of the measurement-device independent quantum key distribution
with phase randomized general sources", Sci. Rep. \textbf{4}, 4612 (2014)

\bibitem{Honji} T. Honjo, K. Inoue, and H. Takahashi, "Differential-phase-shift quantum key distribution experiment with a planar light-wave circuit Mach–Zehnder interferometer", Opt. Lett. \textbf{29}, 2797 (2004).

\bibitem{Azuma} K. Azuma, Weighted sums of certain dependent random variables. Tohoku Math. J. \textbf{19}, 357 (1967).

\bibitem{Decoy3} X. B. Wang, "Beating the photon-number-splitting attack in practical quantum cryptography", Phys. Rev. Lett. \textbf{94}, 230503 (2005).

\bibitem{Decoy2} H. K. Lo, X. F. Ma, K. Chen, "Decoy state quantum key distribution," Phys. Rev. Lett. \textbf{94}, 230504 (2005).

\bibitem{RT3} V. Zapatero1, Á. Navarrete, K. Tamaki and M. Curty, "Security of quantum key distribution with intensity correlations", preprint arXiv:2105.11165v1 (2021).

\end{thebibliography}
\end{document}